\documentclass[a4paper]{jpconf}
\usepackage[pdftex]{graphicx} 

\usepackage[hang,small,bf]{caption}
\usepackage[subrefformat=parens]{subcaption}
\begin{document}
\title{MPGD simulation in negative-ion gas for direction-sensitive dark matter searches}

\author{Hirohisa Ishiura$^1$, Rob Veenhof$^2$, Kentaro Miuchi$^1$, Ikeda Tomonori$^1$}

\address{$^1$Department of Physics, Kobe University, Rokkodaicho, Nada Hyogo 657-8501, Japan}
\address{$^2$MEPhI Kashirskoe shosse, 31 115409, Russia, Moscow and RD51 CERN }

\ead{ishiura@stu.kobe-u.ac.jp}

\begin{abstract}
    Negative-ion time projection chambers (NITPCs) provide the absolute z-position from the arrival-times' difference of several species of negative-ions. This is a strong background reduction method for rare event searches, such as direct dark mater searches. The characteristics of micro pattern gaseous detectors (MPGDs) in negative-ion gas were studied by both experiments and simulations to develop an NITPC for dark matter search experiments. An MPGD simulation method for the negative-ion gas with Garfield++, which has not yet been established, was developed. We present herein the MPGD simulation results in negative-ion gas, $\mathrm{SF_{6}}$ gas, together with the experimental results, such as gain, energy resolution, and signal creation from the MPGD for future NITPC optimization.
\end{abstract}

\section{Introduction}
Many pieces of evidence support the existence of dark matter in the universe. One of the most attracting candidates is weakly interacting massive particles (WIMPs). Three methods can be used to search for WIMPs: direct, indirect and collider experiments. Direct detection aims to catch the signal of nuclear recoil by the WIMPs. Many direct dark matter search experiments have been performed in the world. 
Among the variations of direct detection experiments, directional dark matter searches are said to provide a ''smoking gun'' evidence of the galactic dark matter\ \cite{TANIMORI2004241}. 
Several directional dark matter search experiments utilizing low-pressure gaseous time projection chambers (TPCs), such as DRIFT\ \cite{BATTAT201765}, MIMAC\ \cite{Santos:2013oua}, and NEWAGE\ \cite{Nakamura:PTEP2015}, are being performed.
As is the case for other direct dark matter search experiments, background-reduction is one of the most important R\&D items for the gaseous TPC detectors. Large volume detectors are also required because the expected event rate is very low. However, electron diffusion during drift is large compared to its track length in the conventional TPCs, in which the electrons are drifted by the electric field; thus constructing a large TPC (e.g., 1\ $\mathrm{m^{3}}$ scale) is very challenging.

A negative-ion TPC (NITPC) is a gaseous TPC with negative-ion gas, or gas with a large electron affinity gas. This type of TPC, in which negative-ions are drifted instead of the electrons, has been studied for the high position resolution applications because the ion diffusion is smaller than that of the electrons\ \cite{MARTOFF2000355}.
The DRIFT group pioneered the development of NITPCs for direct dark matter search and succeeded in operating a larger volume TPC. 
The ''minority carrier,'' which makes the z-fiducialization possible\ \cite{Battat:2014van}\ \cite{PHAN2015}.
After the nuclear recoil and ionizations of gas molecules, several species of negative-ion, called main and minority charges, are created in the NITPCs. These ion species have different drift velocities; hence, the absolute z-position of the incident position can be known from the arrival times. This feature enables detectors to cut radioactive decay events from the drift and readout planes. 
First NITPC with $\mathrm{CS_2}$ and a few percent $\mathrm{O_{2}}$ mixture was demonstrated \cite{BATTAT201765}. $\mathrm{SF_6}$ was also recently identified as a good negative-ion gas candidate for NITPC \cite{PHAN2015}. $\mathrm{SF_6}$ is a safer, non-flammable and non-toxic gas compared to the  $\mathrm{CS_{2} + O_{2}}$ mixture.

NITPC is a relatively new technology; thus, no method has yet been established to simulate micro pattern gaseous detectors (MPGDs) in negative-ion gas with Garfield++ \cite{Garfield:url}, which is a MPGD simulation toolkit. 
The NITPC simulation method and their results together with the related experimental results are described herein to establish the MPGD simulation method for negative-ion gas.

\section{Measurements}
\subsection{Measurement Setup}
\begin{figure}[]
    \centering
    \includegraphics[width=8cm]{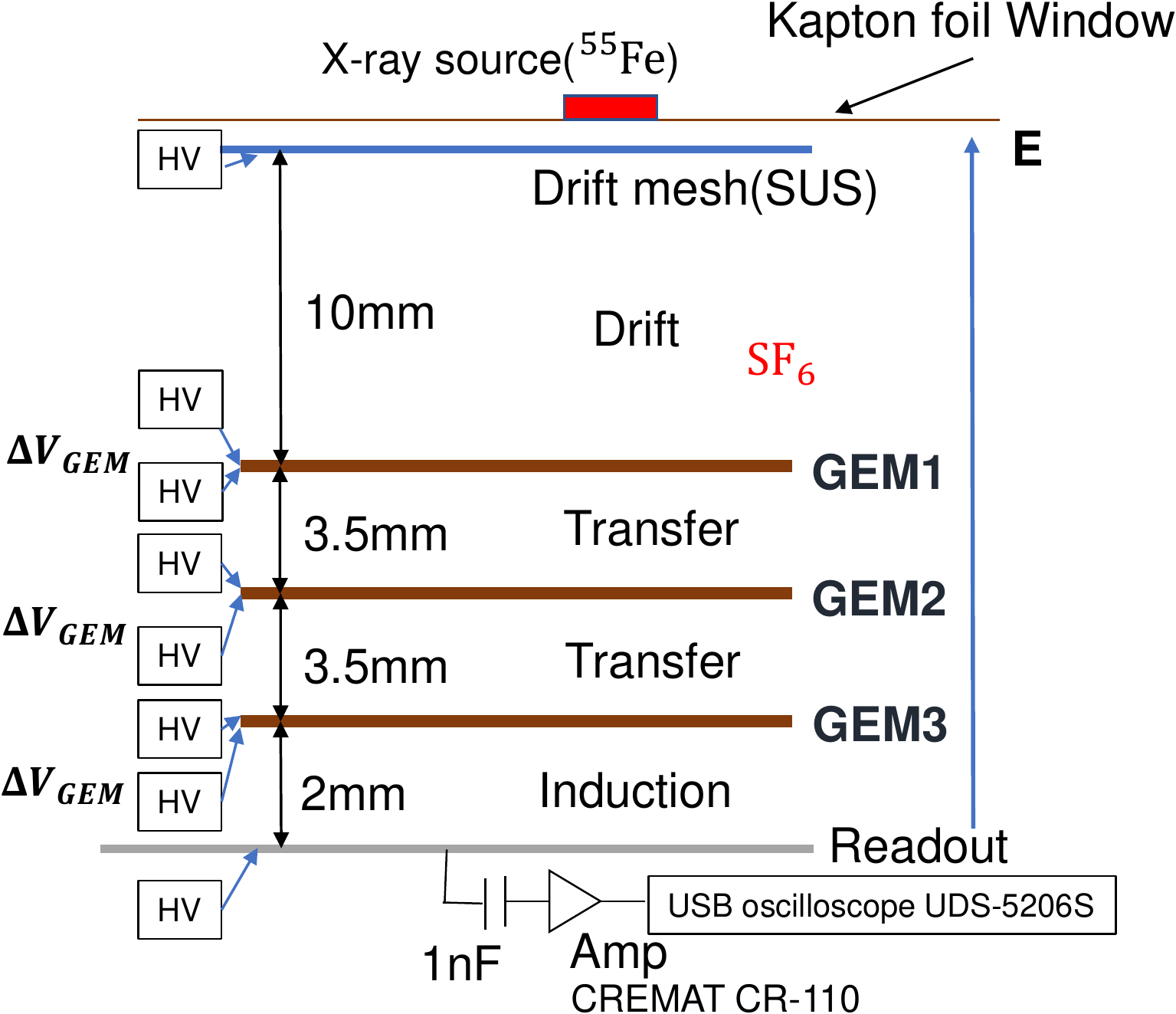}
    \caption{Gain measurement setup}
    \label{fig:exp_setup}
\end{figure}

Several fundamental properties of the MPGD performance in negative-ion gas were measured as a comparison target of the simulation.
Fig.\ref{fig:exp_setup} schematically shows the setup used to evaluate the performance of GEMs\ \cite{SAULI1997531} in $\mathrm{SF_6}$. Two or three GEMs (100\ $\mathrm{\mu m}$ thickness, liquid crystal polymer, made by Scienergy) were used. The difference between these setups was only the number of GEMs. The drift mesh, made of stainless steel, was set on the top, and GEM1 (three-GEM case) or GEM2 (double-GEM case) was located 10\ mm below the drift mesh. The transfer gap, which is the distance between the GEMs, was 3.5\ mm, while the induction gap, which is the distance between GEM3 and the readout plane, was 2\ mm. One-dimensional 24\ strips with a pitch of 400\ $\mu$m were used as the readout electrodes. The signals from these strips were grouped to one channel, decoupled with a 1\ nF capacitor, and read by the charge sensitive amplifier, Cremat CR-110 (Gain:\ 1.4\ V/pC, $\tau$\ =\ 140\ $\mu$s). The waveforms were obtained using a USB oscilloscope UDS-5206S. Pure $\mathrm{SF_6}$ gas with various low pressures of 60--120\ Torr were used. The gas gains were measured with different applied voltages and pressures. 
\subsection{Measurement result}
\begin{figure}
    \begin{minipage}[b]{0.48\linewidth}
        \centering
        \includegraphics[keepaspectratio, scale=0.4]{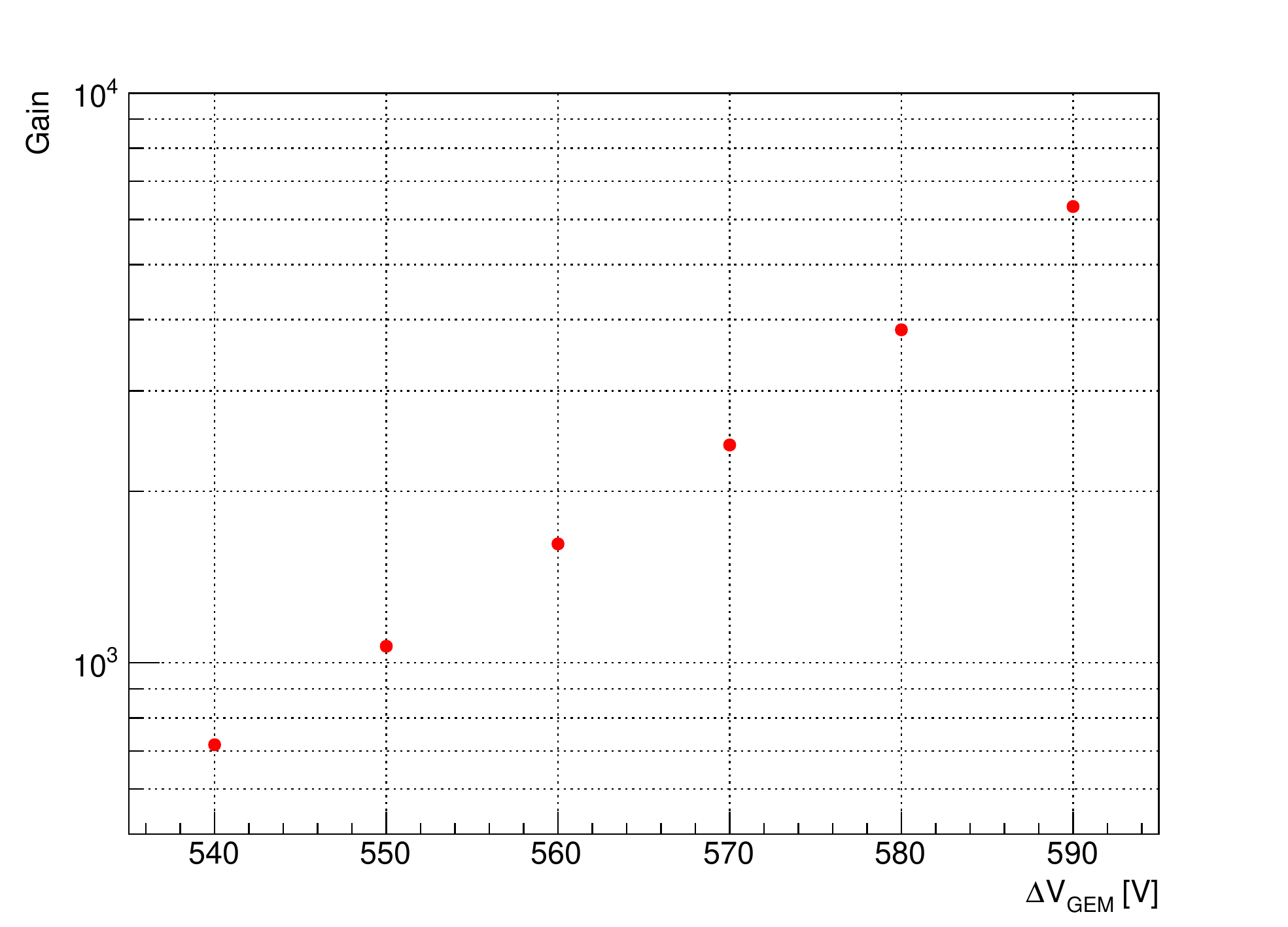}
        \subcaption{Double GEM gas gain at 100\ Torr pure $\mathrm{SF_{6}}$}
        \label{fig:doubleGEM_gain} 
    \end{minipage}
    \begin{minipage}[b]{0.45\linewidth}
        \centering
        \includegraphics[keepaspectratio, scale=0.4]{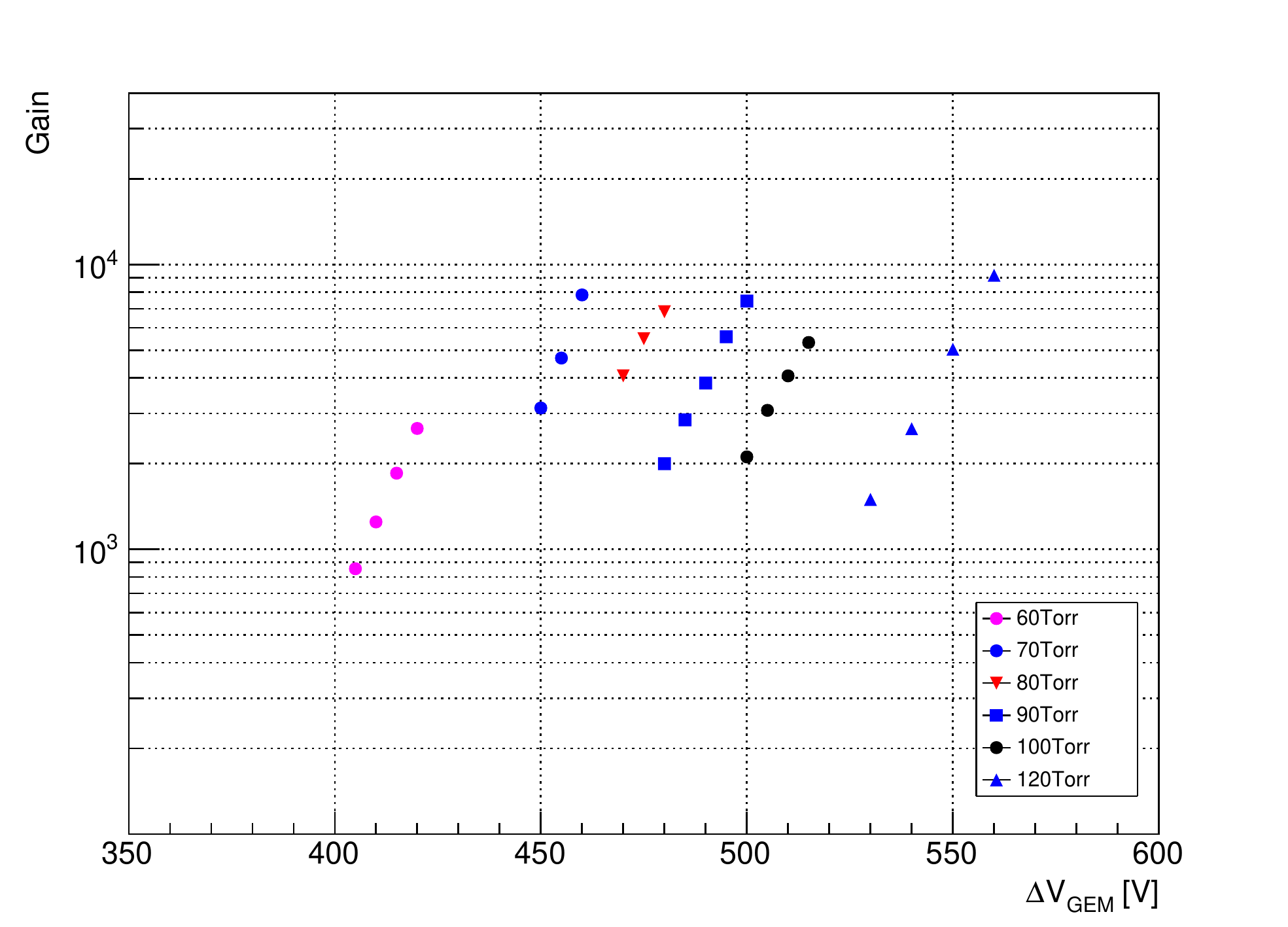}
        \subcaption{Triple GEM gas gain in pure $\mathrm{SF_{6}}$}
        \label{fig:tripleGEM_gain} 
    \end{minipage}
    \caption{Double and Triple GEM gas gain curve}
    \label{fig:GEM_gain}
\end{figure}

\begin{figure}
    \begin{minipage}[b]{0.48\linewidth}
        \centering
        \includegraphics[keepaspectratio, scale=0.4]{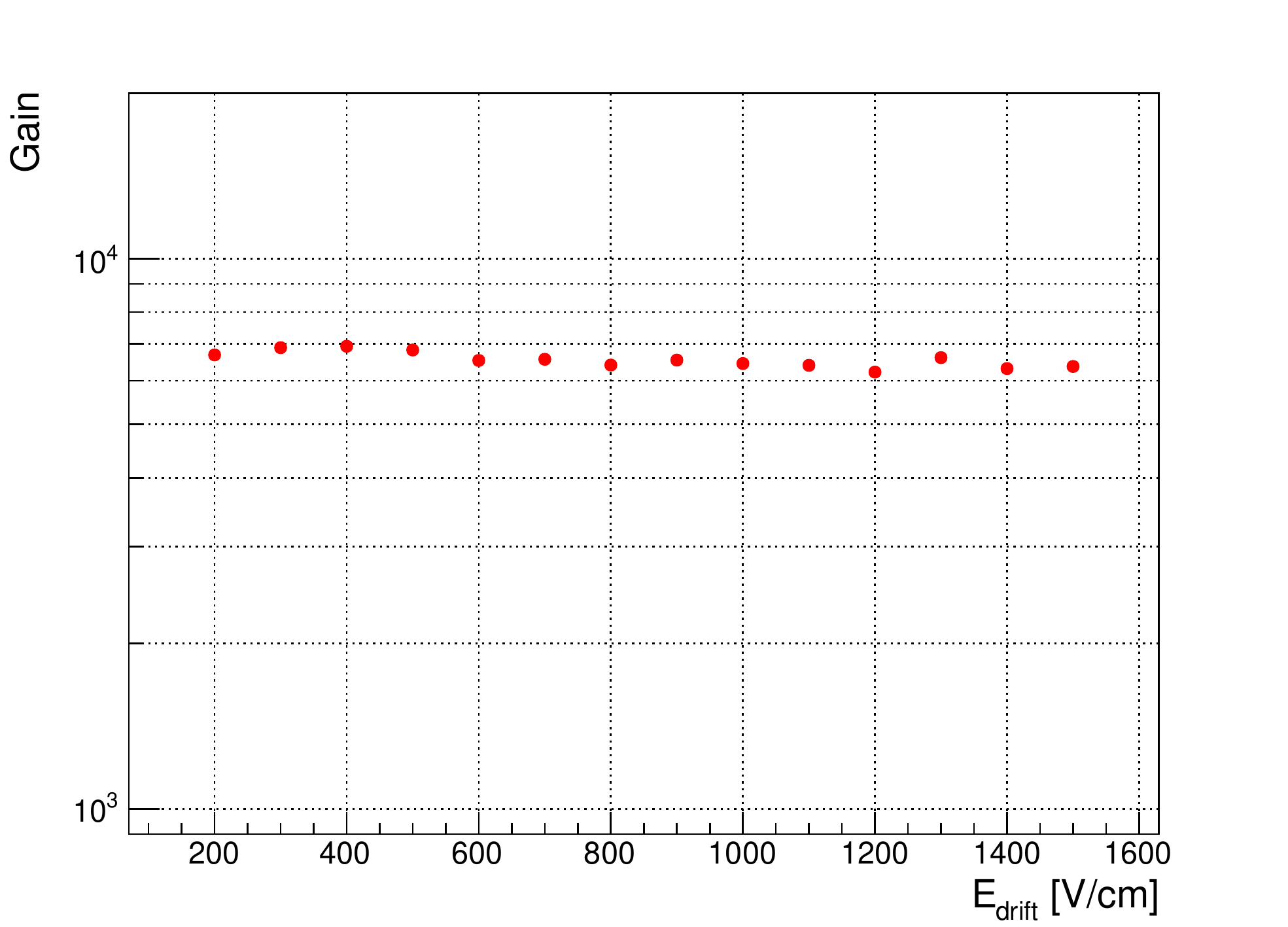}
        \subcaption{Gas gain dependence on $E_{drift}$. Measurement of the conditions are $\mathrm{SF_{6}}$ at 100\ Torr,$E_{transfer}=2.9\ \mathrm{kV/cm}$ and $E_{induction}=2.5\ \mathrm{kV/cm}$}
        \label{fig:drift_dep_result} 
    \end{minipage}
    \begin{minipage}[b]{0.48\hsize}
        \centering
        \includegraphics[keepaspectratio, scale=0.4]{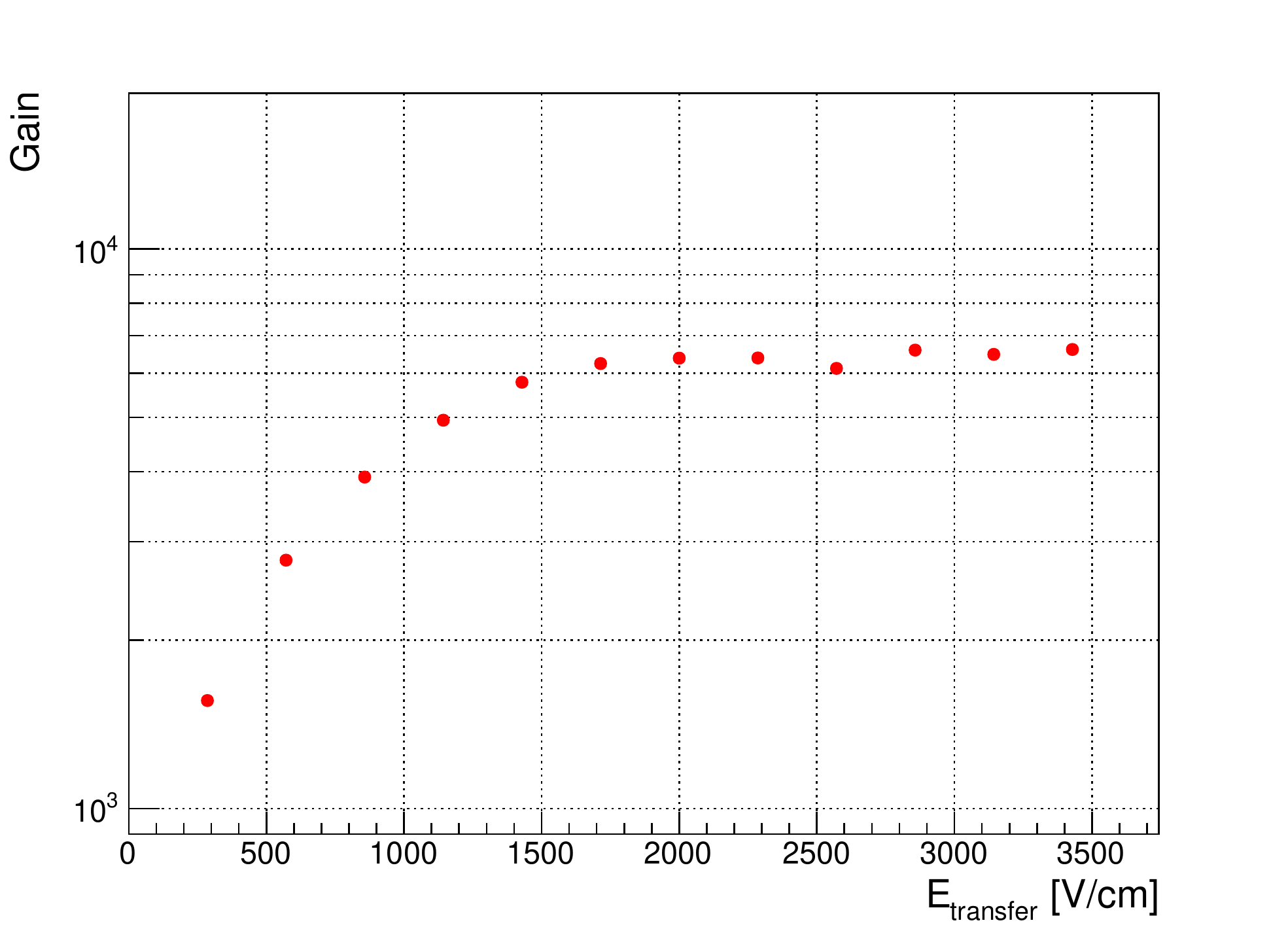}
        \subcaption{Gas gain dependence on $E_{transfer}$. Measurement of the conditions are $\mathrm{SF_{6}}$ at 100\ Torr,$E_{drift}=1.0\ \mathrm{kV/cm}$ and $E_{induction}=2.5\ \mathrm{kV/cm}$}
        \label{fig:transfer_dep_result} 
    \end{minipage}\\
    \centering
    \begin{minipage}[b]{0.48\linewidth}
        \includegraphics[keepaspectratio, scale=0.4]{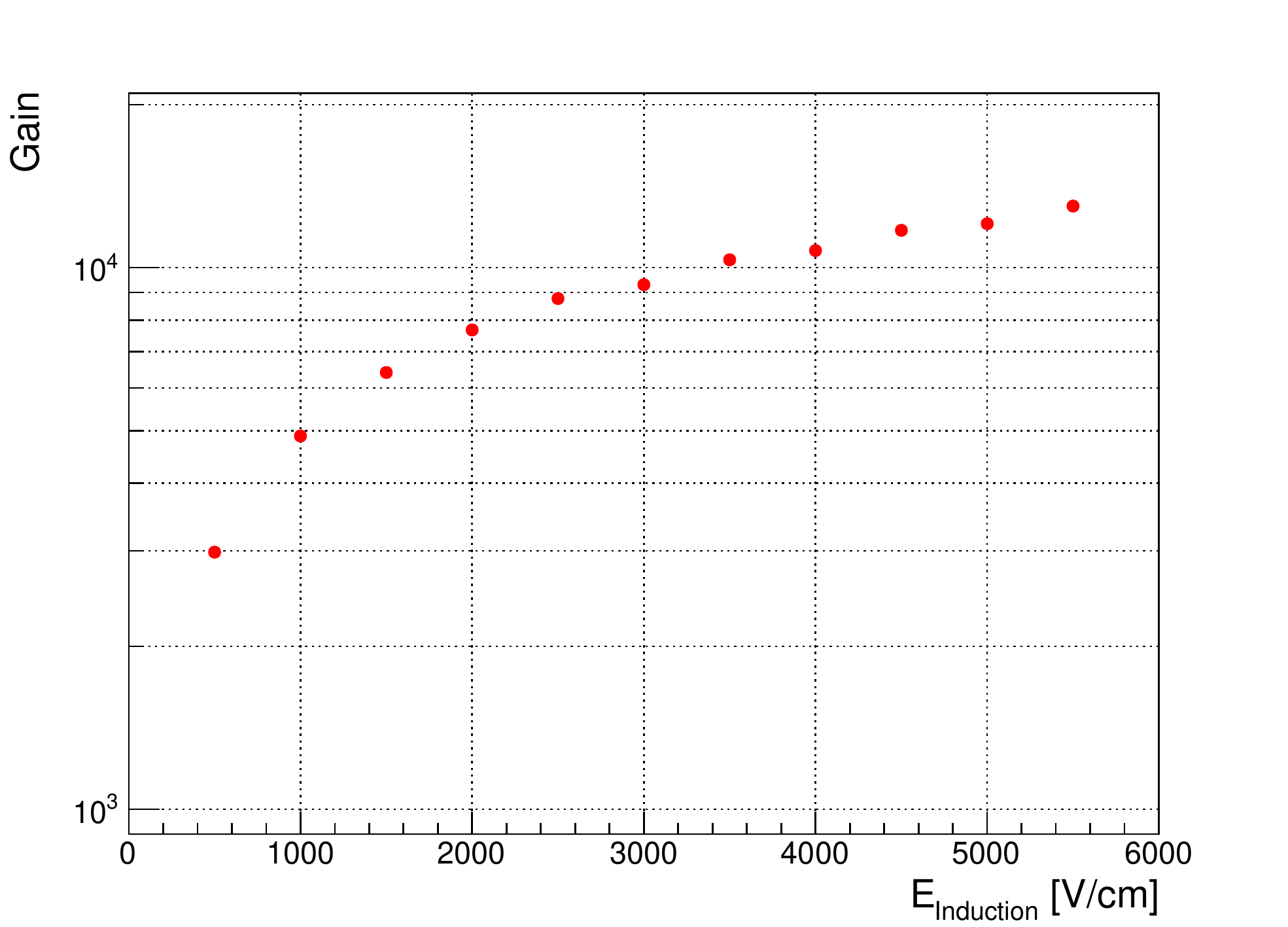}
        \subcaption{Gas gain dependence on $E_{induction}$. Measurement of the conditions are $\mathrm{SF_{6}}$ at 100\ Torr,$E_{drift}=1.0\ \mathrm{kV/cm}$ and $E_{transfer}=2.9\ \mathrm{kV/cm}$}%
        \label{fig:induction_dep} 
    \end{minipage}
    \caption{Gas gain dependence on the electric fields in $\mathrm{SF_{6}}$ at 100\ Torr}
    \label{fig:dep_results}
\end{figure}

Fig.\ref{fig:GEM_gain} shows the measurement results.
Gas gains of up to 6,000 were obtained with $\mathrm{SF_6}$ at 100\ Torr in the double-GEM measurement (Fig.\ref{fig:doubleGEM_gain}).
Gas gains of up to 10,000 for 60 to 120\ Torr were obtained in the triple GEM measurement (Fig.\ref{fig:tripleGEM_gain}).
Figs.\ref{fig:dep_results}a--c show the gas gain dependence on the drift, the transfer and the induction, respectively. 
No dependence on the drift electric field was observed, as shown in Fig.\ref{fig:drift_dep_result}.
Some dependence on the transfer and induction electric fields was observed as shown in Fig.\ref{fig:transfer_dep_result} and Fig.\ref{fig:induction_dep}. In both cases, the gains increased at the lower electric field region and reached plateaus at the higher electric field region. In addition, the induction-dependence measurement result showed a slight gain rise after the plateau, indicating that some electron multiplication may occur at the induction gap.

\begin{figure}[]
    \centering
    \includegraphics[width=.6\textwidth]{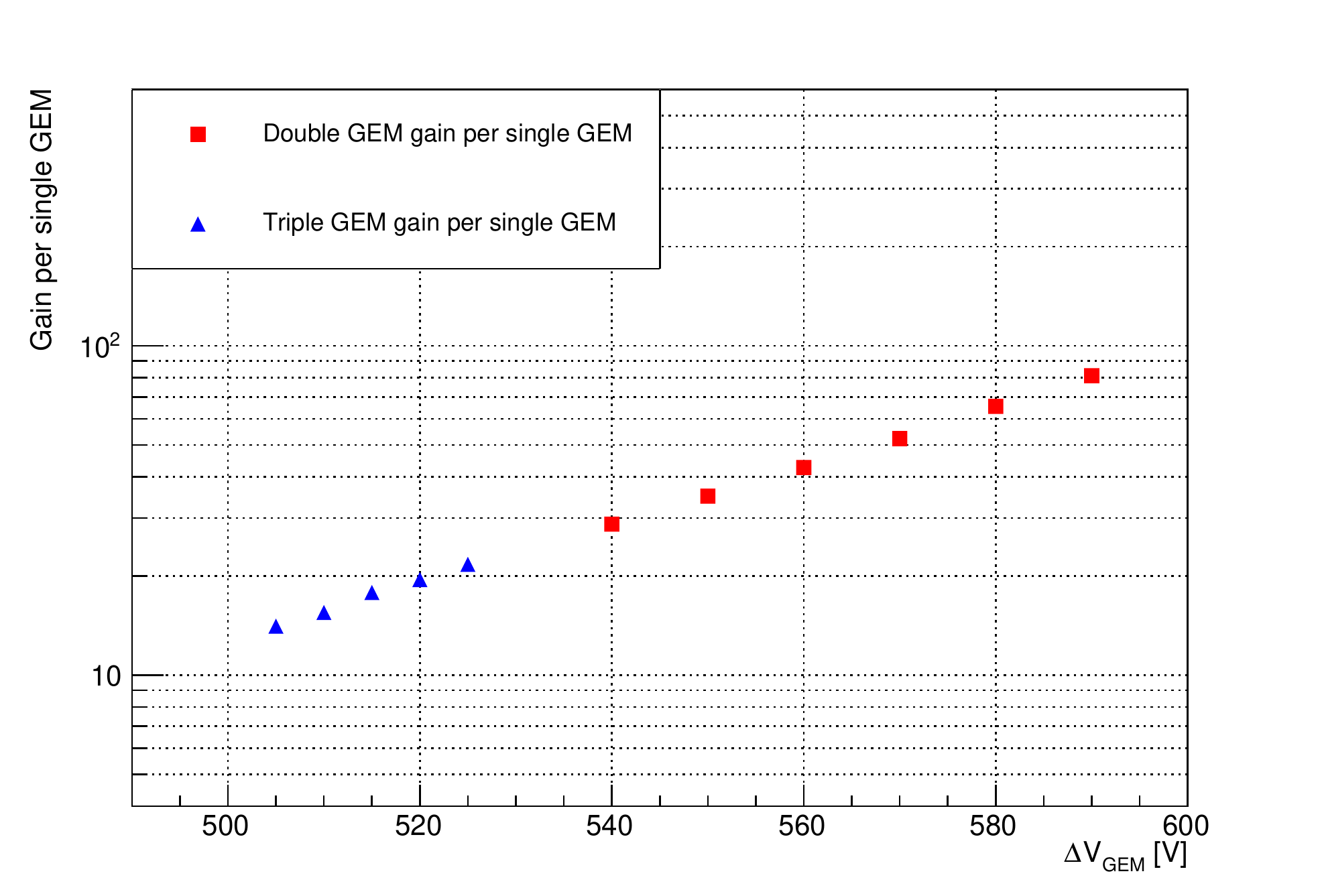}
    \caption{Gas gain per single GEM at 100\ Torr $\mathrm{SF_{6}}$}
    \label{fig:gain_per_single} 
\end{figure}

The gas gains relevant to the single GEM were calculated from the double and triple-GEM results by assuming that the total gain was a product of two or three GEM gains. 
Fig.\ref{fig:gain_per_single} shows the result. The single GEM gains in the case of double and three GEM were on the same gain curve.
This result indicates that there is no significant charge loss in the transfer region between the GEMs. 

    

\section{MPGD simulation}
The MPGD simulation study for the negative-ion gas is indispensable for the optimization of the design and operation. 
Garfield++\ \cite{Garfield:url} and Magboltz\ \cite{Magboltz:url} version 9.01 were used herein. 
Garfield++ is a simulation toolkit for the gaseous detector study written in C++. It imports detector geometry and electric field from the external finite element method software's output and uses the Monte Carlo microscopic and macroscopic methods for the charge propagation.
Magboltz calculates the transport parameters of the electrons in given gas, electric field, and magnetic field from the cross-section data. These parameters were passed to Garfied++ and used for the gas avalanche calculation.

\subsection{Simulation method}
The cross-section data for the reactions between the electron and the negative-ion gas, such as $\mathrm{SF_6}$, were implemented in Magboltz. However no code was implemented for the negative-ion transportation and the electron detachment.
The negative-ion transportation process and two electron detachment models were introduced herein and implemented in the Garfield++ customization code. 
The whole process is presented as follows.
\begin{enumerate}
    \item Negative-ions drift in the electric field and reach the MPGD region. 
    \item The electron is detached from the negative-ion gas molecule by the high electric field\ (electron detachment). 
    \item An avalanche starts from the detached electron.
    \item The electron-ion pairs create the signal.
\end{enumerate}

As a simulation setup, a single 100\ $\mathrm{\mu m}$ thick GEM, which has 140\ $\mathrm{\mu}$m holes pitch and 70\ $\mathrm{\mu}$m hole diameter, was used. The gas was pure $\mathrm{SF_{6}}$\ 100\ Torr and only the main charge $\mathrm{SF_{6}^{-}}$ was considered in the drift process\ (i).
The electron detachment cross-section of $\mathrm{SF_{6}^{-}}$ was used for the electron detachment process\ (ii) and two models, namely the cross-section and threshold models, were introduced and used to simulate the detachment process. The electron~/~$\mathrm{SF_{6}}$ cross-section which Magboltz has was used for the avalanche process\ (iii). 


\subsection{cross-section model} 
\begin{figure}
    \centering
    \begin{minipage}[b]{0.48\linewidth}
        \centering
        \includegraphics[keepaspectratio, scale=0.5]{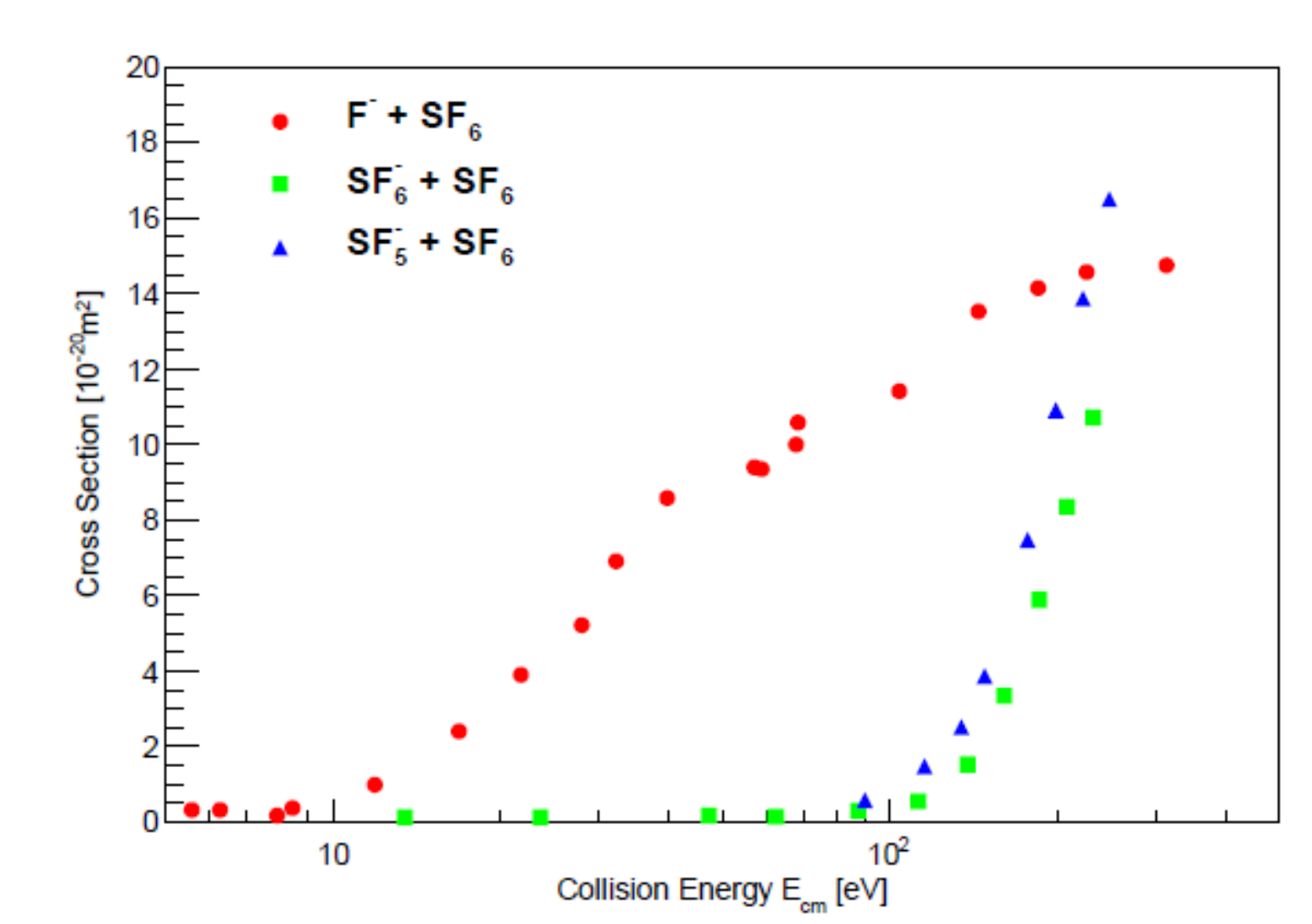}
        \subcaption{Direct detachment, Eq.(\ref{eq:sf6_direct_reaction1})(green square), Eq..(\ref{eq:sf6_direct_reaction2})(blue triangle) and Eq.(\ref{eq:sf6_indirect_reaction3})(red circle)}
        \label{fig:detach_direct} 
    \end{minipage}
    \begin{minipage}[b]{0.48\linewidth}
        \centering
        \includegraphics[keepaspectratio, scale=0.5]{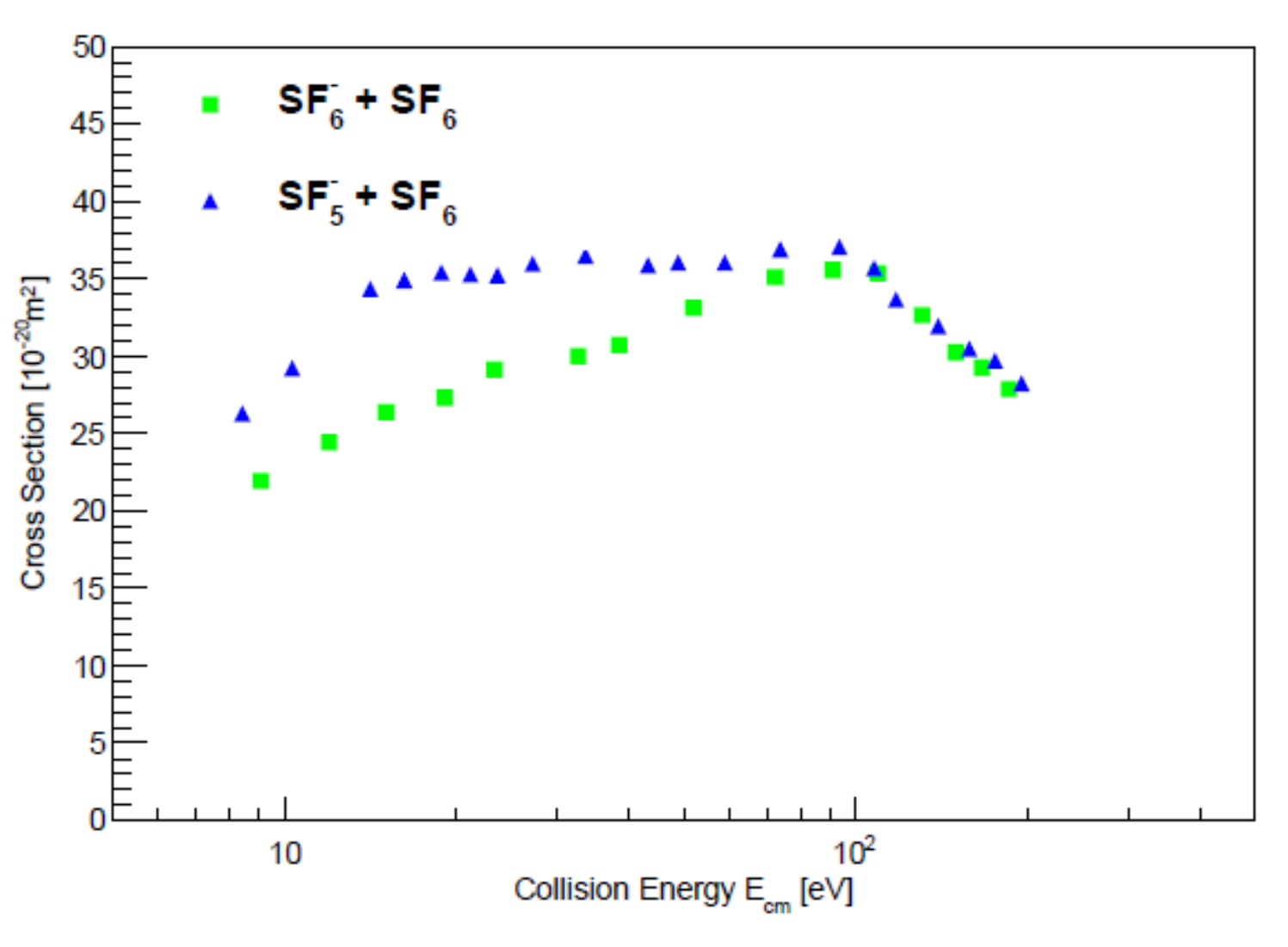}
        \subcaption{Fluorine ion ($\mathrm{F^{-}}$) creation reaction, Eq.(\ref{eq:sf6_indirect_reaction1})(green triangle) and Eq.(\ref{eq:sf6_indirect_reaction2})(blue triangle)}
        \label{fig:detach_indirect} 
    \end{minipage}
    \caption{Direct electron detachment and fluorine ion ($\mathrm{F^{-}}$) creation reactions \cite{YWANG}}
    \label{fig:detachment_reaction_crosssection}
\end{figure}

The electron detachment probability in the cross-section model was calculated using the cross-section of the relevant chemical processes. Two possible detachment processes, namely direct(Eq.(\ref{eq:sf6_direct_reaction1}) or (\ref{eq:sf6_direct_reaction2})) and indirect one (Eq.(\ref{eq:sf6_indirect_reaction1})+(\ref{eq:sf6_indirect_reaction3})~or~(\ref{eq:sf6_indirect_reaction2})+(\ref{eq:sf6_indirect_reaction3})), were considered as possible processes.
The cross-sections of these reactions were reported in \cite{YWANG}.
The cross-section of Eq.(\ref{eq:sf6_direct_reaction1}) and (\ref{eq:sf6_direct_reaction2}) increased at 100\ eV (Fig.\ref{fig:detach_direct}) while that of Eq.(\ref{eq:sf6_indirect_reaction3}) 
increased at 10\ eV. This is ten times lower energy than what Eq.(\ref{eq:sf6_direct_reaction1}),\ (\ref{eq:sf6_direct_reaction2}) showed at 100\ eV. 

Fig.\ref{fig:detach_indirect} indicates that the Eq.(\ref{eq:sf6_indirect_reaction1}) and (\ref{eq:sf6_indirect_reaction2}) reactions have larger cross-sections below 10\ eV compared to Eq.(\ref{eq:sf6_direct_reaction1}) and (\ref{eq:sf6_direct_reaction2}). In addition, $\mathrm{SF_{6}^{-}}$ has larger mass than $\mathrm{F^{-}}$ so $\mathrm{SF_{6}^{-}}$ is difficult to be accelerated and have sufficient energy for the reactions. %
Therefore the detachment rate can be assumed to be determined by Eq.(\ref{eq:sf6_indirect_reaction3}), such that the cross-section of Eq.(\ref{eq:sf6_indirect_reaction3}) can be used for the detachment calculation.

\begin{eqnarray}
    \label{eq:sf6_direct_reaction1}
    SF_{6}^{-} + SF_{6} & \rightarrow & e^{-} + SF_{6} + SF_{6} \\
    \label{eq:sf6_direct_reaction2}
    SF_{5}^{-} + SF_{6} & \rightarrow & e^{-} + SF_{5} + SF_{6} 
\end{eqnarray}
\begin{eqnarray}
    \label{eq:sf6_indirect_reaction1}
    SF_{6}^{-} + SF_{6} & \rightarrow & F^{-} + SF_{5} +SF_{6} \\
    \label{eq:sf6_indirect_reaction2}
    SF_{5}^{-} + SF_{6} & \rightarrow & F^{-} + SF_{4} +SF_{6} \\
    \label{eq:sf6_indirect_reaction3}
    F^{-} + SF_{6} & \rightarrow & e^{-} + F +SF_{6}
\end{eqnarray}
In the Eq.(\ref{eq:sf6_indirect_reaction1}) reaction, the $\mathrm{SF_{6}^{-}}$ energy can be calculated using the $\mathrm{SF_{6}^{-}}$ mobility in $\mathrm{SF_{6}}$\ \cite{Benhenni2012} and the fluorine ion $\mathrm{F^{-}}$ is expected to have energy up to approximately 2\ eV by a simple kinematics calculation.
The $\mathrm{F^{-}/SF_{6}}$ collision reaction has a cross-section of $\mathrm{\sigma=10^{-15}-10^{-14}\ cm^2}$\ \cite{Benhenni2012} in $\mathrm{SF_{6}}$ 100\ Torr at the corresponding energy range, and the number of density is $\mathrm{n = 3.5\times 10^{18}\ cm^{-3}}$. Thus the mean-free-path of this reaction can be calculated to be in an order of $\mathrm{10^{-6}\ m}$. 
The energy of the $\mathrm{F^{-}}$ ion would obtain from the electric field 
can be calculated along its path.
The energy obtained in the GEM hole, where typical electric field is  approximately 50\ kV/cm and above, was 10\ eV if the $\mathrm{F^{-}/SF_{6}}$ collision mean-free-path $\lambda$ is an order of $\mathrm{10^{-6}\ m}$, and this energy is close to Eq.(\ref{eq:sf6_indirect_reaction3}) reaction's threshold. 
Given the $\mathrm{F^{-}}$ energy, the mean-free-path of the electron detachment reaction through the indirect reaction, $\mathrm{\lambda}$, was thus known based on the cross-section shown in Fig.\ref{fig:detach_direct}\ \cite{YWANG}.


The probability ($p$) of the electron detachment from $\mathrm{F^{-}}$ within 1\ $\mathrm{\mu m}$, the same as the step used in this simulation, was calculated as Eq.(\ref{eq:detach_prob})
\begin{eqnarray}
    p = 1 - \exp(- 1 \mu m / \lambda)
    \label{eq:detach_prob}
\end{eqnarray}
This detachment probabilities are shown with the blue triangles in Fig.\ref{fig:detach_prob}. The interpolated probability values were used in the simulation.



\subsection{Threshold model}
The other model, the threshold model, is a more phenomenological one. In this model, the negative-ion releases the electron when the electric field is more than a threshold value. In this simulation, the negative-ion releases an electron in the high electric field larger than 45\ kV/cm, where the avalanches are first observed in the corresponding experiments. This threshold value was known from the electric field map of the GEM geometry when the gas avalanche starts in the experiments. The red line in Fig.\ref{fig:detach_prob} shows the probability function.

\subsection{Simulation result}
\begin{figure}[]
    \centering
    \includegraphics[width=10cm]{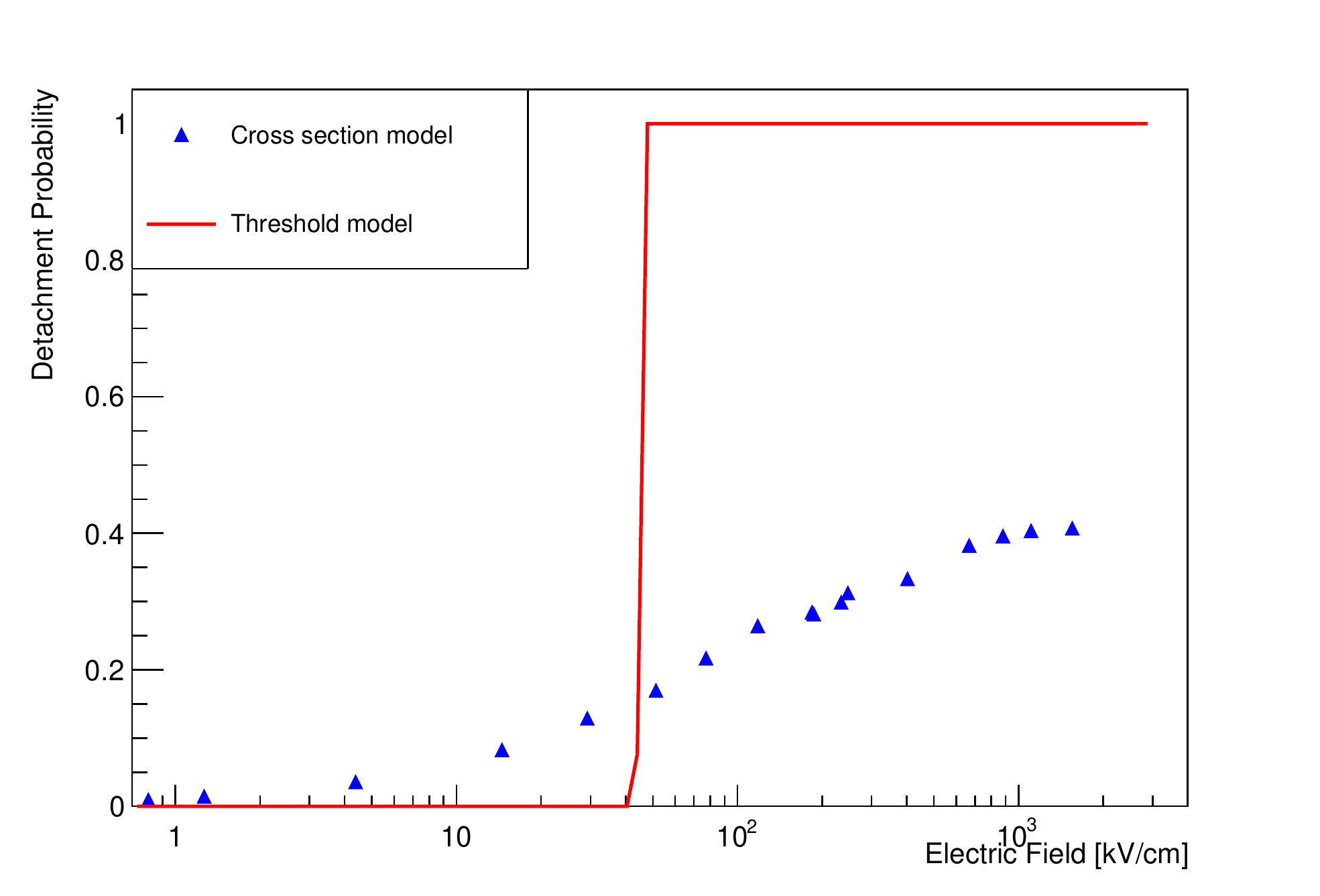}
    \caption{Detachment probability dependence on the electric field. The blue marker denotes the cross-section model, while red line denotes the threshold model probability function.}
    \label{fig:detach_prob}
\end{figure}
\begin{figure}[]
    \centering
    \includegraphics[width=12cm]{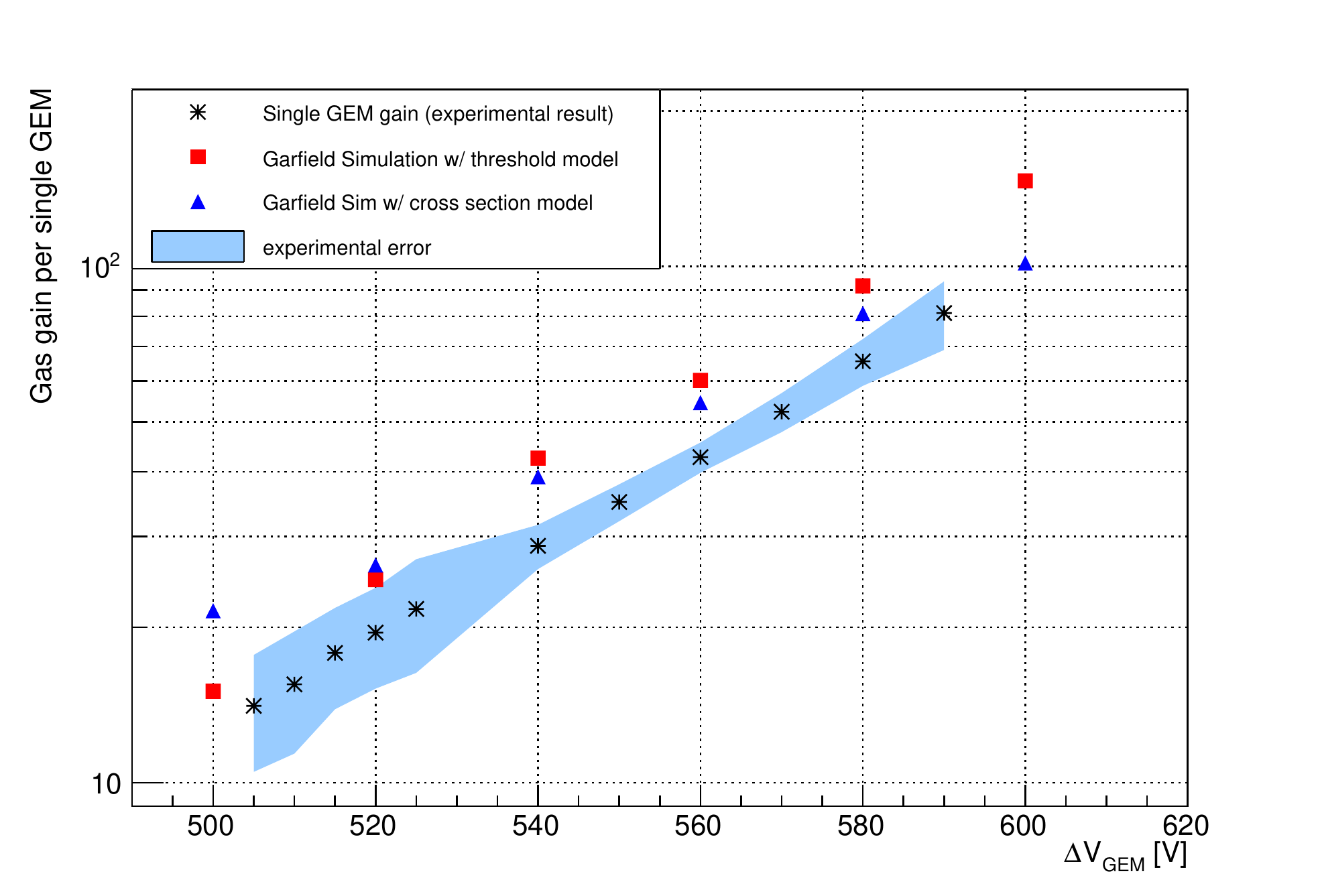}
    \caption{Gas gain curve (pure\ $\mathrm{SF_6}$\ 100\ Torr). The cross mark denotes the measurement result and red square denotes threshold model, and blue triangle denotes the cross-section model.}
    \label{fig:final_result}
\end{figure}
The gas avalanche simulations were performed with the detachment models described above. Fig.\ref{fig:final_result} shows the simulation results. The blue triangles and the red squares show the results of the cross-section model and the threshold models, respectively.
The result of the measurements and its error are also depicted as the asterisk and the light blue region, respectively. It should be noted, as a primary outcome of this study, the absolute gas gain was reproduced by the simulation within factor 2 difference. This result can be claimed as a great success of this first study. 
The main difference between the cross-section model and threshold models was the slope of the gain curve. In other words, the cross-section model had a larger gain than threshold model at low electric field but did not at higher electric field. These results can be explained by the detachment probability difference, shown in Fig.\ref{fig:detach_prob}, such that at low electric field cross-section model had a larger detachment probability and, on the contrary, at higher electric field threshold model had.
A comparison of the experimental and simulation results indicated that the threshold model produced the experimental result more than the cross-section model in terms of the slope of the gain curve. 
Since the cross-section model is a more first principle method, tuning the cross-section method would be a important step to generalize the outcome of this study.

Studies for other MPGDs, such as MicroMegas and $\mathrm{\mu}$-PIC in negative-ion gas are also important future works.
It is also important to compare the gas gain dependences on the transfer and induction electric field. Studies on other parameters, such as the energy  resolution, signal formation, and signal collection, would help to validate this method. An optimization work are also planned for dark matter search experiment.

\section{Conclusion}
In this work, the first MPGD simulation study with $\mathrm{SF_{6}}$, one of negative-ion gases, was achieved with Garfield++ and Magboltz. Two detachment models, cross-section and threshold model, were introduced as the electron detachment process. In both models, the absolute gas gain was reproduced by the simulation within factor 2 difference. In the terms of the slope of the gain curve, the threshold model reproduced the experiments result more than cross-section one. 
\section*{Acknowledgement}
This work was supported by KAKENHI Grants-in-Aids (15K21747, 26104005, 16H02189 and 19H05806), Grant-in-Aid for JSPS Fellows (17J03537 and 19J20376).
\section*{References}

\bibliographystyle{iopart-num}
\bibliography{ref}

\end{document}